\documentclass{DISproc}
\usepackage[utf8x]{inputenc}
\usepackage[english]{babel}

\if@mathematic
   \def\vec#1{\ensuremath{\mathchoice
                     {\mbox{\boldmath$\displaystyle\mathbf{#1}$}}
                     {\mbox{\boldmath$\textstyle\mathbf{#1}$}}
                     {\mbox{\boldmath$\scriptstyle\mathbf{#1}$}}
                     {\mbox{\boldmath$\scriptscriptstyle\mathbf{#1}$}}}}
\else
   \def\vec#1{\ensuremath{\mathchoice
                     {\mbox{\boldmath$\displaystyle#1$}}
                     {\mbox{\boldmath$\textstyle#1$}}
                     {\mbox{\boldmath$\scriptstyle#1$}}
                     {\mbox{\boldmath$\scriptscriptstyle#1$}}}}

\begin{document}
\title{Dijet Production in QCD and ${\cal N}=4$ SYM}

\author{{\slshape Grigorios Chachamis$^1$, José Daniel Madrigal$^2$, Agustín Sabio Vera$^2$}\\[1ex]
$^1$Paul Scherrer Institut, CH-5232 Villigen PSI, Switzerland\\
$^2$Instituto de Física Teórica UAM/CSIC,U. Autónoma de Madrid, E-28049 Madrid, Spain.}

\contribID{131}

\doi  

\maketitle

\begin{abstract}
We investigate dijet production at large rapidity separation in QCD and ${\cal N}=4$ SYM, showing that both 
theories give similar predictions for observables only sensitive to conformal properties of the scattering such as 
ratios of azimuthal angle correlations. Renormalization prescriptions are important in this comparison. 
\end{abstract}

\section{Introduction}

In spite of real-world QCD being neither supersymmetric nor conformal invariant, it shares many features with ${\cal N}=4$ SYM. Both theories are identical when loops of quarks and scalars do not appear, i.e. at tree-level, and in the leading $\ln s$ term of the high-energy limit, where the scattering is dominated by exchange of a gluon ladder in the $t$-channel. The infrared structure of both theories is similar at the level of soft divergences.  It is interesting to look for regimes where QCD and ${\cal N}=4$ give similar predictions. With this target in mind we will study both theories in the Regge limit $s\gg-t$, with $s$ and $t$ the usual Mandelstam invariants. 

\section{Ratios of Azimuthal Correlations in High-Energy Dijets}

\begin{figure}[htb]
  \centering
  \includegraphics[width=0.5\textwidth]{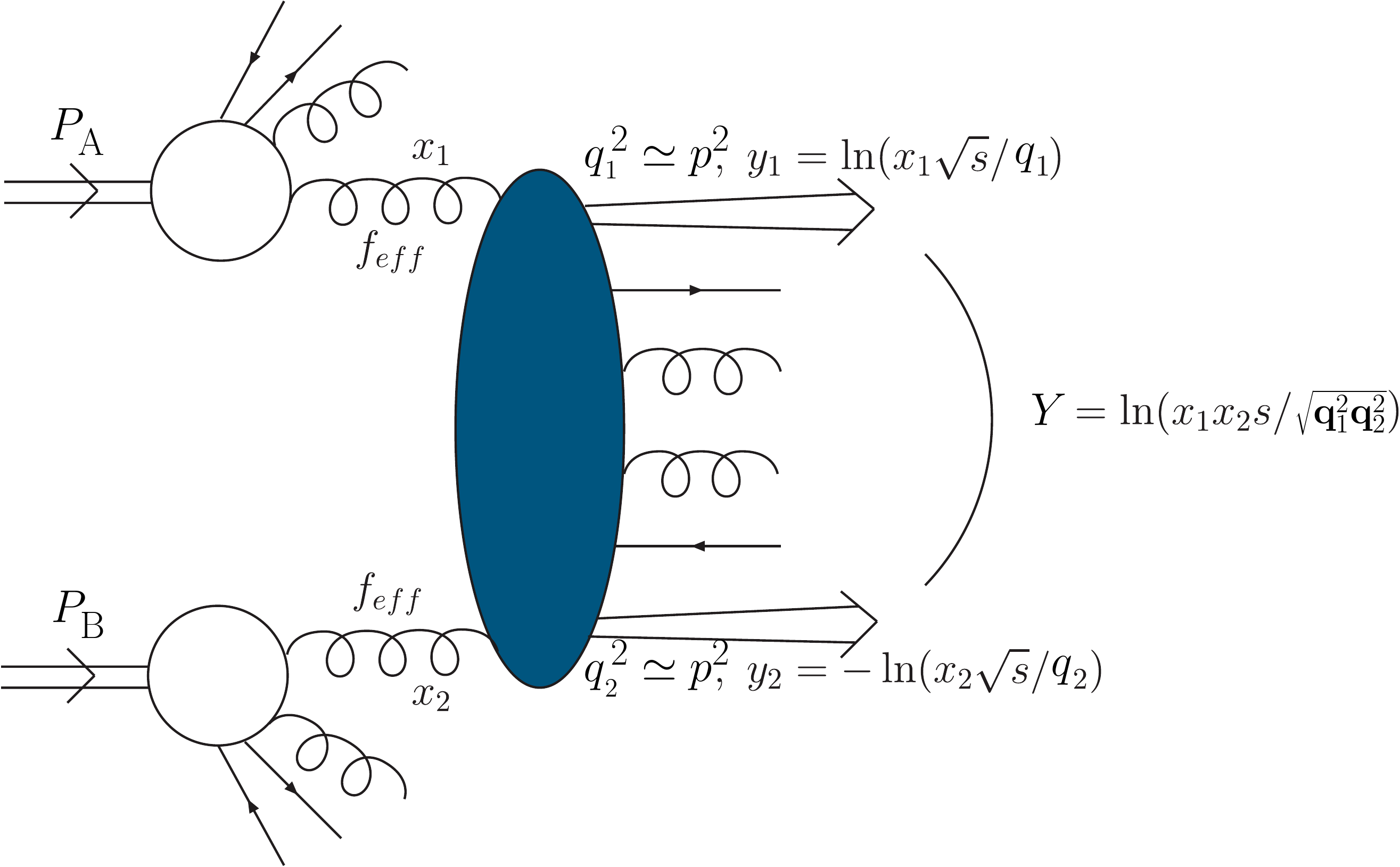}
  \caption{Mueller-Navelet Jets.}
  \label{Dijet}
\end{figure}

To define the observables of interest, we consider the kinematic configuration of Fig. \ref{Dijet}, the well-known Mueller-Navelet jets \cite{Mueller:1986ey}. Two forward jets with similar transverse scales $q_{1,2}^2\simeq p^2$ are tagged at a large rapidity separation $Y=\ln\frac{x_1x_2 s}{\sqrt{\vec{q}_1^2\vec{q}_2^2}}$ and relative azimuthal angle $\phi=\vartheta_1-\vartheta_2$, where $x_{1,2}$ are the fractions of longitudinal momenta of the parent hadrons carried by the jet\footnote{Partons are here identified with jets, as we consider for simplicity the jet vertex only to leading order \cite{Angioni:2011wj}.}. Such a configuration is particularly suited to unveil QCD dynamics in the high-energy limit, specifically through the study of dijet azimuthal correlation \cite{DelDuca:1993mn,Goerlich}.

The large value of $Y$ calls for a resummation of high-energy logarithms of the form $(\alpha_s\ln(s/q^2))^n$. This resummation is performed in the BFKL approach \cite{Lipatov:1976zz}, and the differential cross section for dijet production at the parton level\footnote{Dependences on PDFs cancel in our observables, allowing for sound comparison between QCD and ${\cal N}=4$.} is given by

\begin{equation}\label{1}
\frac{d\hat{\sigma}}{d^2\vec{q}_1d^2\vec{q}^2}=\frac{\pi^2\bar{\alpha}_s^2}{2}\frac{f(\vec{q}_1,\vec{q}_2,Y)}{\vec{q}_1^2\vec{q}_2^2};\quad f(\vec{q}_1,\vec{q}_2,Y)=\int\frac{d\omega}{2\pi i}e^{\omega Y} f(\vec{q}_1,\vec{q}_2,\omega),\quad \bar{\alpha}_s\equiv\frac{\alpha_s N_c}{\pi},
\end{equation}
in terms of the Mellin transform of the solution to the BFKL equation

\begin{equation}
\omega f(\vec{q}_1^2,\vec{q}_2^2,\omega)=\delta^2(\vec{q}_1^2-\vec{q}_2^2)+\int d^2\vec{\kappa}\,{\cal K}_{\rm NLL}(\vec{q}_1,\vec{\kappa})f(\vec{\kappa},\vec{q}_2,\omega).
\end{equation}

In the leading log approximation, this equation enjoys conformal SL(2,$\mathbb{C})$ invariance in the plane transverse to the colliding partons \cite{Lipatov:1985uk}, and the integral kernel ${\cal K}$ is diagonalised by the eigenfunctions $\psi_{n,\nu}=\frac{1}{2\sqrt{\pi}}(\vec{q}^2)^{i\nu-1/2}e^{in\vartheta}$. The discrete quantum number $n=0,1,2\cdots$ controls the azimuthal behaviour in the transverse plane and corresponds to a conformal spin, since it carries a representation of SL(2,$\mathbb{C})$. It turns out \cite{Vera:2006un} that the asympotic intercepts of the kernel corresponding to $n\ge 1$ are very similar at LL and NLL, not having the weak convergence problem of the $n=0$ case ---corresponding to the pomeron intercept--- which needs an all-orders collinear resummation~\cite{Salam:1998tj}. This motivates us to look for observables insensitive to the conformal spin $n=0$.\\

We consider the Fourier expansion of the differential cross section \eqref{1} in the azimuthal angle
\begin{equation}
\begin{gathered}
\frac{d\hat{\sigma}(\bar{\alpha}_s,Y,p^2)}{d\phi}=\frac{\pi\bar{\alpha}_s^2}{4p^2}\sum_{n=-\infty}^{\infty}e^{in\phi}{\cal C}_n(Y);\\
{\cal C}_n^{\rm QCD}(Y)=\int_{-\infty}^\infty \frac{d\nu}{2\pi}\tfrac{\exp\left[{{\bar \alpha}_s \left(p^2\right) Y \left(\chi_0\left(\left|n\right|,\nu\right)+{\bar \alpha}_s  \left(p^2\right) \left(\chi_1\left(\left|n\right|,\nu\right)-\tfrac{\beta_0}{8 N_c} \tfrac{\chi_0\left(\left|n\right|,\nu\right)}{\left(\frac{1}{4}+\nu^2\right)}\right)\right)}\right]}{1/4+\nu^2}.
\end{gathered}
\end{equation}

The label $n$ for each of the moments is the conformal spin. The functions $\chi_{0,1}$ in the definition of the coefficients ${\cal C}_n$ are the building blocks of the NLL BFKL kernel \cite{Fadin:1998py}. A similar formula for ${\cal C}_n$ holds in the supersymmetric case with $\beta_0=0$ and $\bar{\alpha}_s$ replaced by the corresponding 't Hooft coupling $a$, see \cite{Angioni:2011wj} for details. Different observables can be built out of the coefficients ${\cal C}_n$. The total (averaged over $\phi$) cross-section is $\hat{\sigma}=\tfrac{\pi^3\bar{\alpha}_s^2}{2p^2}{\cal C}_0$, while contributions of higher conformal spins are projected in the moments $\langle \cos(n\phi)\rangle={\cal C}_n/{\cal C}_0$. The dependence on $n=0$ cancels when constructing the ratios ${\cal R}_{m,n}=\frac{\langle \cos (m\phi)\rangle}{\langle \cos (n\phi)\rangle}=\frac{{\cal C}_m}{{\cal C}_n}$. These ratios ${\cal R}_{m,n}$ have a good perturbative convergence, and are clean observables to test the properties of the Regge limit, in particular conformal invariance, without interference of collinear contributions. They were computed in \cite{Angioni:2011wj} in QCD and ${\cal N}=4$ SYM for different renormalisation prescriptions.
 
 \section{Comparing QCD and ${\cal N}=4$ SYM}
 
\begin{figure}[htb]
  \centering
  \includegraphics[width=0.85\textwidth]{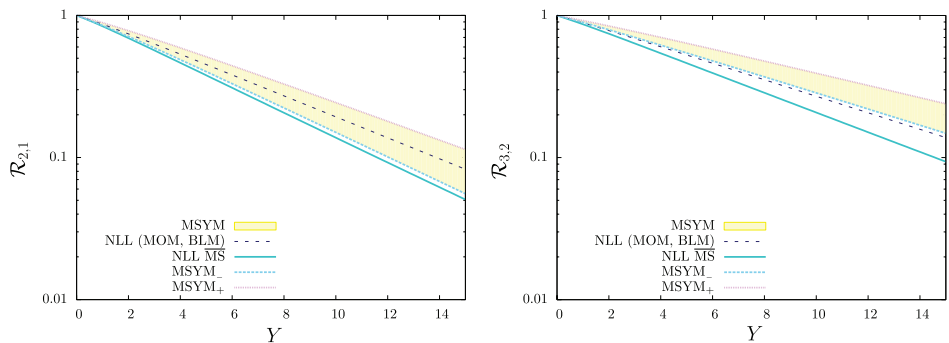}
  \caption{Evolution of ratios ${\cal R}_{2,1}$ (left) and ${\cal R}_{3,2}$ (right) with jet rapidity separation in QCD and
${\cal N}=4$ SYM for different renormalisation schemes ($\overline{\rm MS}$ vs BLM).}
  \label{ratio}
\end{figure}

 The results found in \cite{Angioni:2011wj} are summarised in Fig. \ref{ratio}. ${\cal N}=4$ results are showed in a yellow band since the non-running coupling was allowed to take values in a given range. One can see that the predictions of the two theories for ${\cal R}_{m,n}$ lie very close, in particular when choosing the Brodsky-Lepage-Mackenzie (BLM) procedure \cite{Brodsky:1982gc}. The BLM procedure effectively resums the effects of a non-zero $\beta$-function and has a  conformal behaviour like the one exhibited by ${\cal N}=4$ theory. In \cite{Brodsky:2001ye} it was shown to give a more sensible result than $\overline{\rm MS}$ scheme for the pomeron intercept. The appearance of an unnaturally high BLM scale for the coupling in this case is relaxed when studying observables insensitive to $n=0$ \cite{Chachamis:2011tg}. It is interesting to note that only when considering clean and perturbatively convergent observables for the high-energy limit, the BLM prescription is systematically closer than the other renormalization schemes 
 to the supersymmetric result.
 
{\footnotesize Research supported by E. Comission [LHCPhenoNet
(PITN-GA-2010-264564)] \& C. Madrid (HEPHACOS
ESP-1473).} 

{\raggedright
\begin{footnotesize}



\end{footnotesize}
}


\end{document}